\documentclass[eqsecnum,floats,aps,footinbib,prd,twocolumn,floatfix,showpacs]{revtex4-1} 

\usepackage{latexsym,pifont,textcomp,setspace,xspace}
\usepackage{amsmath,amstext,amssymb,color,graphicx}
\usepackage{bm}
\usepackage[utf8]{inputenc}
\usepackage{soul} %stikeout
\def\d{{\mathrm{d}}}

\usepackage{amsmath}
\usepackage{amssymb}
\usepackage{indentfirst}
\everymath{\displaystyle}
\usepackage{graphicx}% Include figure files
\usepackage{setspace}

\usepackage{color}

\usepackage[english]{babel}
\definecolor{purple}{rgb}{1,0,1}

\allowdisplaybreaks
\begin{document}
%-----------------------------------------------------------------------------------------------------
\title{Tolman-like temperature gradients in stationary spacetimes}
\author{Jessica Santiago and Matt Visser\vspace{3pt}}
\affiliation{School of Mathematics and Statistics,
Victoria University of Wellington; \\
PO Box 600, Wellington 6140, New Zealand.}
%-----------------------------------------------------------------------------------------------------
\begin{abstract}
%-----------------------------------------------------------------------------------------------------
\noindent
It is (or should be) well known that specification of a heat bath requires \emph{both} a temperature and a 4-velocity, the rest frame of the heat bath. In \emph{static} spacetimes there is a very natural and \emph{unique} candidate for the 4-velocity of the heat bath, the normalized timelike Killing vector. However in stationary non-static spacetimes the situation is considerably more subtle, and several different ``natural'' 4-velocity fields suitable for characterizing the rest frame of a heat bath can be defined --- thus Buchdahl's 1949 analysis for the Tolman temperature gradient in a stationary spacetime is only part of the story. In particular, the heat bath most suitable for describing the Hawking radiation from a rotating black hole is best described in terms of a gradient flow normal to the spacelike hypersurfaces, not in terms of Killing vectors.

\medskip\noindent
{\sc Date:} 6 July  2018, LaTeX-ed \today.

\end{abstract}
%-----------------------------------------------------------------------------------------------------
\pacs{04.20.-q;  04.40.-b; 05.20.-y; 05.70.-a}
%-----------------------------------------------------------------------------------------------------
\maketitle
%-----------------------------------------------------------------------------------------------------
\def\eref#1{(\ref{#1})}
\parindent0pt
\parskip7pt
\def\L{{\mathcal{L}}}

%-----------------------------------------------------------------------------------------------------
\section{Introduction}
%-----------------------------------------------------------------------------------------------------
In 1930, Tolman~\cite{tolman:1930} and later Tolman and Ehrenfest~\cite{ehrenfest:1930}, discovered the existence of relativistic temperature gradients for fluids in thermal equilibrium in static spacetimes. (See also the related references~\cite{tolman:1928,tolman:1933a,tolman:1933b}.) In more modern language, the locally measured temperature is typically presented as:
\begin{equation}
T(x) = {T_0 \over||K||},
\end{equation}
where $||K||$ is the norm of the \emph{static} Killing vector which is parallel to the fluid 4-velocity, and $T_0$ is a position-independent constant, which represents the physical temperature at the zero gravitational potential hypersurface.
(For recent discussions see references~\cite{santiago:2018,grf:2018}. For somewhat older reviews on relativistic thermodynamics see references~\cite{Israel-Stewart:1980, Israel:1981, Israel:1986}.) 
Herein we shall argue that the standard presentation requires some important caveats and extensions.

In 1949 Buchdahl~\cite{buchdahl:1949} formally extended Tolman's result to fluids in stationary spacetimes following the timelike Killing vector 
\begin{equation}
K^a = (\partial_t)^a = (1,0,0,0)^a,
\end{equation}
with a result that looks superficially identical to Tolman's result,  but with some additional subtleties that we shall explore and extend in the current article.

One key point is this: Temperature is certainly a scalar, but defining a heat bath also requires you to specify the 4-velocity of the heat bath. In his 1928--1933 papers, Tolman was always dealing with static spacetimes, where the notion of a preferred 4-velocity exists and to pick the static observer was a very natural choice. This can most easily be seen by putting the metric into its preferred block diagonal form,
\begin{equation}
\label{e:static}
ds^2 =  g_{tt} dt^2 + g_{ij} \, dx^i dx^j,
\end{equation}
in which the ``Killing flow'' is defined to be the \emph{unique} naturally defined 4-velocity:
\begin{equation}
\label{e:killing-flow}
V^a = \hat K^a =  {K^a\over||K||} =  {(1,0,0,0)^a\over \sqrt{-g_{tt}}}.
\end{equation}
(As discussed below, this Killing flow is compatible with thermal equilibrium.)
One can then re-write Tolman's result for static spacetimes as
\begin{equation}
T(x) ={T_0\over||K||} = {T_0\over\sqrt{-g_{tt}}}.
\end{equation}

However, this result needs to be modified
for stationary non-static spacetimes. We shall argue that Buchdahl's 1949 result~\cite{buchdahl:1949}, known and used up to today, is actually incomplete and valid only for a very specific class of 4-velocities. Choosing an ADM-like decomposition for the metric,
\begin{equation}
\label{e:stationary}
ds^2 = - N^2 dt^2 + h_{ij} \,(dx^i-v^i \,dt)\, (dx^j-v^j\, dt), 
\end{equation}
with inverse
\begin{equation}
g^{ab} = \left[\begin{array}{c|c} 
- 1/N^2 & -v^j/N^2\\ \hline -v^i/N^2 & h^{ij} - v^i v^j/N^2
\end{array}\right],
\end{equation}
there is now no unique naturally defined 4-velocity.

One possible option is to nevertheless keep using the Killing flow, though even the Killing flow is not unique in stationary spacetimes with axial symmetry (such as the physically important Kerr and  Kerr--Newman black holes~\cite{Kerr, Kerr-intro, Kerr-book}).
We shall have more to say on this point below; where we consider arbitrary (constant coefficient) linear combinations of the time translation and axial Killing vectors.

Another appealing option is to consider the ``normal flow'' (which is orthogonal to the constant time slices):
\begin{equation}
\label{e:normal}
\hat N_a = - {\nabla_a t \over ||\nabla t||} = N\; (-1,0,0,0)_a.
\end{equation}
In static spacetimes the normal flow and Killing flow can be made to coincide, but not otherwise.

Collectively, these observations indicate that some care and delicacy should be invoked when extending the discussion of Tolman temperature gradients to 
stationary spacetimes --- this is the entire import of the discussion below.

%-----------------------------------------------------------------------------------------------------
\section{Photon gas}
%-----------------------------------------------------------------------------------------------------
\label{s:photon}
%-----------------------------------------------------------------------------------------------------

Start by observing that a photon gas in internal equilibrium satisfies the equation of state
\begin{equation}
\rho = 3 p = a \,T^4.
\label{e:eos}
\end{equation}
Here the $\rho=3p$ condition comes from the fact that photons have zero rest mass, 
while $a$ is the radiation constant coming from the Stefan--Boltzmann law. Now consider the relativistic Euler equation for a perfect fluid
\begin{equation}
(\rho+p) \, A_a =  - (\delta_a{}^b + V_a V^b)\, \nabla_b p.
\label{e:euler}
\end{equation}
For a photon gas this first simplifies to
\begin{equation}
A_a =  - (\delta_a{}^b + V_a V^b)\, \nabla_b \ln T.
\label{e:euler2}
\end{equation}
But, as an absolute minimum condition, internal thermal equilibrium implies 
\begin{equation}
V^b \, \nabla_b T =0.
\end{equation}
(In thermal equilibrium the temperature should be time-independent as one moves with the fluid.) So the Euler equation further reduces to
\begin{equation}
A_a =  - \nabla_a \ln T. 
\label{e:euler3}
\end{equation}
This now intimately connects thermal gradients with the 4-acceleration of the photon fluid. 

However, as shown by Tolman and Ehrenfest \cite{ehrenfest:1930}, and recently discussed in a modern reinterpretation~\cite{santiago:2018,grf:2018} of Maxwell's two-column argument~\cite{maxwell:1868,maxwell:1902}, the temperature gradient must not depend on the substance, nor on the state of matter. Therefore this result, equation (\ref{e:euler3}), is automatically extended to arbitrary systems in internal thermal equilibrium.

Specifically, for any photon gas in free-fall we have $A=0$, and so $T(x)$ is actually a position-independent constant, as expected. Tolman temperature gradients are zero for any fluid following a geodesic path.

In counterpoint, if the heat bath is accelerating, (that is, the 4-acceleration is non-zero), then expanding around some fiducial point $x_0^a$, to lowest order we have
\begin{equation}
T(x) = T(x_0) \; \left\{ 1 + A_a (x^a-x_0^a) + O([\Delta x]^2) \right\}.
\end{equation}
Therefore, for any accelerating thermal bath, we \emph{do} expect temperature gradients in thermal equilibrium.

Making this statement even more clear: Equation \eqref{e:euler3} tells us the relation between the fluid's 4-acceleration and its temperature gradient, regardless of whether the spacetime is Minkowski, or Schwarzschild,  or Kerr--Newman. The spacetime can be flat, curved, stationary, static, whatever --- if the 4-acceleration of the fluid (assumed to obey the relativistic Euler equation and to be in internal equilibrium) is given, the temperature gradient can be obtained.

%-----------------------------------------------------------------------------------------------------
\section{Tolman 1930: Killing flow}
%-----------------------------------------------------------------------------------------------------
\vspace{-10pt}
For completeness, let us now see how a simplified derivation of Tolman's result can be obtained. (The derivation is simplified in the sense that this derivation makes it clear that the Einstein equations are absolutely not necessary for obtaining relativistic temperature gradients.)

Consider a static spacetime with the metric presented in the block-diagonal form of equation~\eref{e:static}. It is a standard well-known result that world-lines ``at rest'' are subject to a non-zero 4-acceleration given by
\begin{equation}
A_a = \nabla_a \ln \sqrt{-g_{tt}}.
\end{equation}
(A formal proof of this result will be subsumed into the more general Buchdahl result discussed below.)
Combined with equation~\eref{e:euler3} this immediately leads to the condition $T(x) \sqrt{-g_{tt}} = \hbox{(constant)}$, which is Tolman's key result~\cite{tolman:1930,ehrenfest:1930}. As alluded to above, extending this result to stationary spacetimes requires care and delicacy.

%-----------------------------------------------------------------------------------------------------
\section{Buchdahl 1949: Killing flow}
%-----------------------------------------------------------------------------------------------------

From a modern perspective Buchdahl's 1949 result can be extended as follows:
Suppose we have some arbitrary timelike Killing vector (not necessarily the time-translation Killing vector; neither does it need to be hypersurface orthogonal) in a spacetime which is either static or stationary.
Now assume a fluid following some world-line in this metric. We want to know whether this system will exhibit Tolman-like temperature gradients or not.
If we choose the fluid to follow integral curves of the Killing vector, as in 
\begin{equation}
\label{e:killing-flow}
V^a = \hat K^a =  {K^a\over||K||},
\end{equation}
then the fluid 4-acceleration can be easily computed. 

We start by noting that
\begin{eqnarray}
K^a \nabla_a(g_{bc} K^b K^c) &=& 2 g_{bc}(K^a \nabla_a K^b) K^c 
\nonumber\\
&=& 2 K^a \nabla_{(a} K_{c)} K^c = 0.
\end{eqnarray}
We now compute:
\begin{eqnarray}
A_a &=& V^b \nabla_b V_a =  V^b \nabla_b \left(K_a\over||K||\right) 
= {V^b \nabla_b K_a\over||K||}.\qquad
\end{eqnarray}
Here we have used the fact that $g_{ab}K^aK^b = - ||K||^2$, so  $K^b \nabla_b ||K|| =0$. 
Applying Killing's equation, 
\begin{eqnarray}
A_a &=&  - {V^b \nabla_a K_b\over||K||} = {1\over2} {\nabla_a( ||K||^2)\over ||K||^2}.
\end{eqnarray}
Then
\begin{equation}
\label{e:A-killing}
A_a =  \nabla_a \ln ||K||.
\end{equation}
This purely kinematic result, valid for any Killing flow, really is the key part of Buchdahl's 1949 result; the rest is (in view of Tolman's 1930 analysis) straightforward.

Combined with equation~\eref{e:euler3} this immediately leads to
\begin{equation}
T(x) = {T_0\over||K||}.
\end{equation}
Here $K$ is now \emph{any} timelike Killing vector, \emph{as long as the fluid follows integral curves of that same Killing vector}.

It is then clear how temperature gradients depend on the system's 4-velocity. For a distorted rotating body (without axial symmetry) there will only be one timelike Killing vector, keeping life simple.
But for a stationary axisymmetric spacetime (for example the Kerr or Kerr--Newman spacetimes) there are \emph{two ``fundamental''} Killing vectors --- the time-translation and rotational Killing vectors.
 Any (constant) linear combination of these Killing vectors  is again a Killing vector --- so there are \emph{infinitely many} timelike Killing vectors to choose from, each one with a different norm, resulting in distinct internal temperature gradients.

The physics message here is this: When applying the Tolman temperature gradient argument in stationary spacetimes, even if you restrict attention to Killing flows,  you have to extremely carefully specify the 4-velocity of the particular thermal bath you are interested in. The situation is even more subtle once one considers normal flows.  

%\vspace{-20pt}
%-----------------------------------------------------------------------------------------------------
\section{Equilibrium Normal flow}
%-----------------------------------------------------------------------------------------------------

In a stationary spacetimes the other ``natural'' option is to take the fluid to follow a normal flow, such that  $V^a \propto -g^{ab}\, \nabla_b t$. (See equation \eref{e:normal}.) 

Explicitly,
\begin{equation}
\label{e: normal flow}
V^a = \hat N^a = {(1; v^i)\over N},
\end{equation}
or even 
\begin{equation}
V^a =  -{\nabla^a t \over||\nabla t||};
\qquad
||\nabla t|| = \sqrt{-g^{tt}}= {1\over N}.
\end{equation}
Here the minus sign is introduced to keep $V^a$ future-directed. Besides that, normal flows are automatically vorticity free in the sense that  $\omega = *(V \wedge \d V) = 0$, a commonly occurring condition in many physically interesting situations \cite{CMP,vorticity}.

By construction we have $K^a\nabla_a t = 1$ for the timelike Killing vector. This implies that, for the Lie derivative, we have:
\begin{eqnarray}
\L_K \nabla_a t &=& K^b \nabla_b \nabla_a t+ \nabla_a K^b \nabla_b t = 0.
\end{eqnarray}
That is, $\L_K \nabla_a t =0$, implying  $\L_K ||\nabla_a t|| =0$.
Also, since we want the fluid travelling along the normal flow to be in internal equilibrium, the fluid should see a ``time-independent'' environment. Whence we must demand the two (somewhat nontrivial) compatibility conditions, 
\begin{equation}
\label{e:compatibility1}
V^a\nabla_a N =0;  
\end{equation}
and
\begin{equation}
\label{e:compatibility2}
V^a\nabla_a p = 0.
\end{equation}
The second compatibility condition is actually the natural extension of the previously imposed thermal equilibrium condition $V^a\nabla_a T = 0$, originally applied to a photon gas to obtain \eqref{e:euler3},
but now extended to general fluids.

For such an equilibrium-compatible normal flow, calculating the 4-acceleration is easy but  slightly different from the calculation for a Killing flow:
\begin{eqnarray}
A_a &=& V^b \nabla_b V_a =  -V^b \nabla_b \left(\nabla_a t\over||\nabla t||\right)
=
 -{V^b \nabla_b \nabla_a t\over||\nabla t||}.\qquad
\end{eqnarray}
We cannot now apply Killing's equation, instead we can use $\nabla_b \nabla_a t = \nabla_a \nabla_b t$, so that
\begin{eqnarray}
A_a &=& - {V^b \nabla_a \nabla_b t\over||\nabla t||} 
= -{1\over2} {\nabla_a( ||\nabla t||^2)\over ||\nabla t||^2}.
\end{eqnarray}

\clearpage
In this way, for a normal flow satisfying the compatibility condition \eref{e:compatibility1}, we have the following purely kinematic result:
\begin{eqnarray}
\label{e:A-normal}
A_a &=&   - \nabla_a \ln ||\nabla t||.
\end{eqnarray}
Given equation \eqref{e: normal flow}, this is equivalent to
\begin{eqnarray}
\label{e:A-normal2}
A_a &=&  \nabla_a \ln N.
\end{eqnarray}
Note $V^a A_a =0$.  This is formally somewhat similar to Buchdahl's result for Killing flows, see equation \eref{e:A-killing},
with $||K||\to N$. 

In \emph{static} spacetimes (in block diagonal form) we have $g_{tt} \, g^{tt} = 1$, implying that for the time translation Killing vector 
$||\nabla t|| \, ||K|| =1$. Therefore, for static spacetimes, both Tolman's original computation for 4-acceleration as the normal flow calculation just shown can be made to coincide. For \emph{stationary} spacetimes, on the other hand, they can and typically will be physically different.

Combined with equation~\eref{e:euler3} this immediately leads to
\begin{equation}
\label{e:new}
T(x) = T_0 \; ||\nabla t|| = T_0 \; \sqrt{-g^{tt}} = {T_0\over N}.
\end{equation}
This is the analogue of Buchdahl's 1949 result, but now applied to (equilibrium compatible) normal flows.
Note this is a very different physical setup from the Buchdahl 1949 result~\cite{buchdahl:1949}, even if the final result superficially looks very similar.

%-----------------------------------------------------------------------------------------------------
\section{Example: Free-fall normal flow}
%-----------------------------------------------------------------------------------------------------
Let us now consider several simple illustrative examples.
%-----------------------------------------------------------------------------------------------------
\subsection{Schwarzschild/Reissner--Nordstrom}
%-----------------------------------------------------------------------------------------------------
For either Schwarzschild or Reissner--Nordstrom spacetimes let us choose to use Painleve--Gullstrand coordinates~\cite{painleve,gullstrand,unruh:1981,unexpected,acoustic,LRR}, wherein
\begin{equation}
ds^2 = -dt^2 + \left(dr - \sqrt{2m(r)\over r} \; dt\right)^2 + r^2(d\theta^2+\sin^2\theta\,d\phi^2).
\end{equation}
(The spacetime is in this case static, but not manifestly static, since we have chosen to write the metric in non-diagonal form.)

Consider the normal flow $V^a \propto - g^{ab}\,\nabla_b t$. Then $N=1$, and $||\nabla t|| = 1/N =1$, from which equation \eref{e:A-normal2} implies a zero 4-acceleration. That is, our ``reference fluid'' is in free-fall. Using then equation \eqref{e:euler3}, we obtain that $T(x)=\hbox{(constant)}$.

So we explicitly see that a fluid in a freely falling box (in Schwarzschild or Reissner--Nordstrom spacetime) will \emph{not} exhibit a Tolman temperature gradient (which is exactly what you should expect based on the equivalence principle).
Furthermore, this particular normal flow automatically satisfies the compatibility conditions~\eref{e:compatibility1} and ~\eref{e:compatibility2} \emph{a priori}.

This agrees with our previous general analysis of section \ref{s:photon},
but now we have a completely explicit expression for the 4-velocity of the relevant thermal bath:
\begin{equation}
V^a =  \left( 1; \sqrt{2m(r)/r},0,0\right).
\end{equation}

%-----------------------------------------------------------------------------------------------------
\subsection{Static spherically symmetric spacetimes}
%-----------------------------------------------------------------------------------------------------
Any  static spherically symmetric spacetime can (at least locally) be put in the form
\begin{equation}
ds^2 = -dt^2 + h(r) \left(dr - v(r) \; dt\right)^2 + r^2(d\theta^2+\sin^2\theta\,d\phi^2).
\end{equation}
(The spacetime is static, but not manifestly static, since we have chosen to write the metric in non-diagonal form.)

The normal flow is now
\begin{equation}
V^a =  \left( 1; v(r),0,0\right),
\end{equation}
and is again geodesic.  A freely falling fluid following this trajectory will not see any Tolman temperature gradient.

%-----------------------------------------------------------------------------------------------------
\subsection{Kerr/Kerr--Newman}
%-----------------------------------------------------------------------------------------------------
For the Kerr or Kerr--Newman spacetime let us choose to work in the Doran coordinate system~\cite{Kerr-intro,doran,river}.  
The normal flow, in these Doran coordinates, is
\begin{eqnarray}
\hat N_a &=& - {\nabla_a t} = {(-1;0,0,0)_a}.
\end{eqnarray}
We have $||\nabla t|| = N^{-1} =1$. (See for instance references~\cite{Kerr-intro,doran,river}.) 
From equation \eref{e:A-normal2} this implies $A=0$. That is, our ``reference fluid'' is now in free-fall, obeying the compatibility conditions \eref{e:compatibility1} and \eref{e:compatibility2}, and we again deduce $T(x)=\hbox{(constant)}$.

Thus, again we see that a gas confined in a freely falling box (in Kerr or Kerr--Newman spacetime) will \emph{not} exhibit a Tolman temperature gradient (which is exactly what you should expect based on the equivalence principle).

\vfill

%\vspace{-0.3cm}
%-----------------------------------------------------------------------------------------------------
\section{Example: Temperature gradients for Black Hole spacetimes}
%-----------------------------------------------------------------------------------------------------

\subsection{Some Killing flows}
When dealing with axisymmetric spacetimes in which the geometry is asymptotically flat, the ``natural'' timelike Killing vector is $(1,0,0,0)$, adopting coordinates $(t,r,\theta,\phi)$, and the rotational Killing vector is $(0,0,0,1)$. In this way, vectors of the form $(1,0,0,\Omega)$ will also be timelike Killing vectors for such a spacetime.

Looking at some interesting cases, 
\begin{itemize}
	\item 
	The $\Omega=0$ Killing vector $(1,0,0,0)$ is well behaved at spatial infinity, giving us:
		\begin{equation}
	T(x) = {T_0\over\sqrt{-g_{tt}}} = {T_0\over\sqrt{N^2-h_{ij} v^i v^j}},
	\end{equation}
	where $v^i$ is defined in \eqref{e:stationary}.
	 However, for both Kerr or Kerr--Newman, its norm $||(1,0,0,0)||$ is zero \emph{at the ergosurface} --- not \emph{at the horizon}.  
	
	\item
	For Kerr or Kerr--Newman, setting  $\Omega\to\Omega_H$ the angular velocity of the horizon, the Killing vector $(1,0,0,\Omega_H)$ has a norm $||(1,0,0,\Omega_H)||$, which is zero \emph{at the horizon} --- not \emph{at the ergosurface}.  
	But this Killing vector has the annoying feature that its norm also vanishes in the asymptotic region, near $r\sin\theta \approx1/\Omega_H$. (This is merely an ``annoyance'', not a ``problem'', the same thing happens for a rotating coordinate system in flat Minkowski space.)
	In this situation
	\begin{equation}
	T(x) = {T_0\over\sqrt{N^2-h_{\phi\phi} (v_\phi-\Omega_H)^2}}.
	\end{equation}
	This clearly is a different generalization of Tolman's result.
\end{itemize}
So either the Killing vector is well behaved at spatial infinity, but problematic at the ergo-surface; or it is well-behaved at the horizon but problematic sufficiently far from the axis of rotation. Worse, if we take a generic constant $\Omega$ such that $0\neq \Omega\neq \Omega_H$ then the resulting Killing vector $K^a=(1,0,0,\Omega)$ has null surfaces (and so formally infinite local Tolman temperatures) that correspond neither to the horizons nor to the ergosurfaces. This now leads us to analyze what happens when the flows are not generated by Killing vectors.

\vspace{-10pt}
\subsection{ZAMO normal flow}
\vspace{-10pt}

In the specific case of axial symmetry, the normal flow $V^a \propto - g^{ab} \nabla_b  t$ is often referred to as a ZAMO flow;  the ``Zero Angular Momentum Observer'' flow. Now let us further specialize to Boyer--Lindquist coordinates, where (under mild technical conditions) we can, using $(t,r,\theta,\phi)$ coordinates, block diagonalize the metric into the form~\cite{Wald,MTW}:
\begin{equation}
g_{ab} = \left[\begin{array}{c|cc|c} 
g_{tt}  & 0 & 0& g_{t\phi}\\
\hline
0 & g_{rr} &0 &0\\
0&0&g_{\theta\theta}&0\\
\hline
g_{t\phi}&0&0&g_{\phi\phi}
\end{array}\right].
\end{equation}
The inverse metric is easily computed
\begin{equation}
g^{ab} = 
\left[\begin{array}{c|cc|c} 
g_{\phi\phi}/ g_2  & 0 & 0& - g_{t\phi} /g_2 \\
\hline
0 & 1/g_{rr} &0 &0\\
0&0&1/g_{\theta\theta}&0\\
\hline
-g_{t\phi} /g_2 &0&0&g_{tt} /g_2
\end{array}\right].
\end{equation}
Here $g_2 = g_{tt} \,g_{\phi\phi}-g_{t\phi}^2$, and $\det(g_{ab})= g_2\,g_{rr}\,g_{\theta\theta}$.

Note that $g_{tt}=0$ defines the ergo-surfaces, where the time translation Killing vector $(1;0,0,0)^a$ becomes null. In contrast, horizons are defined by the condition $g^{tt}=\infty$, equivalent to $(g^{tt})^{-1}=0$. If $g_{t\phi}\to 0$, then horizons and ergosurfaces coalesce, but for $g_{t\phi}\neq 0$ they are distinct.

The normal flow, in these Boyer--Lindquist coordinates, is then
\begin{eqnarray}
\hat N_a &=& - {\nabla_a t \over ||\nabla t||} = {(-1;0,0,0)\over \sqrt{-g^{tt}}} = 
 \sqrt{-g_2\over g_{\phi\phi}} \; (-1;0,0,0)
 \nonumber\\
 &=& \sqrt{-g_{tt} +{g_{t\phi}^2\over g_{\phi\phi} }} \;(-1;0,0,0).
\end{eqnarray}
The corresponding flow vector (contravariant vector) is 
\begin{equation}
V^a = \hat N^a =  \sqrt{g_{\phi\phi}\over -g_2 } \; \left(1;0,0,-{g_{t\phi}\over g_{\phi\phi}}\right).
\end{equation}
In terms of the time translation and axial Killing vectors,
(and now defining $\varpi = - g_{t\phi}/g_{tt}$), we have
\begin{equation}
V^a = {[K_T]^a + \varpi [K_\phi]^a\over ||K_T + \varpi K_\phi||}.
\end{equation}
This is not a (normalized) Killing vector, because $\varpi$ is not a constant, it still has $(r,\theta)$ dependence. Indeed we have
\begin{eqnarray}
||K_T + \varpi K_\phi||^2 &=&  - (g_{tt} + 2 \varpi g_{t\phi} +\varpi^2 g_{\phi\phi})
\nonumber\\
&=& - \left(g_{tt} - {g_{t\phi}^2\over g_{\phi\phi}}\right)  
\nonumber\\
&=& - {g_2\over g_{\phi\phi}} =  -{1\over g^{tt}} = N^2.
\end{eqnarray}
This particular normal flow automatically satisfies the compatibility conditions~\eref{e:compatibility1} and \eref{e:compatibility2}. (Because both $N$ and $p$ are functions of $(r,\theta)$ only, whereas the vector $V^a$ lies in the $(t,\phi)$ plane.)

Since this is a special case of a normal flow we still find
\begin{equation}
T(x) = T_0 \, ||\nabla t|| =  {T_0\over N} = T_0 \sqrt{-g^{tt}}.
\end{equation}
So for this particular ZAMO gradient flow, which is definitely not a Killing flow, the redshifted temperature is well behaved from just above the horizon all the way out to spatial infinity. 
This observation is useful for thinking about how to redshift the Hawking temperature for Kerr and Kerr--Newman black holes from the horizon (where the locally measured Hawking temperature diverges) out to spatial infinity (where the locally measured Hawking temperature is finite).

Note that the choice of coordinates (eg, Boyer--Lindquist versus Doran) does not change the physics; rather the choice of coordinates guides one as to choosing some physically appropriate 4-velocity for the heat-bath; and it is this physical choice of 4-velocity for the heat-bath that is responsible for physical differences in the Tolman temperature gradient.

%-----------------------------------------------------------------------------------------------------
\section{Conclusions}
%-----------------------------------------------------------------------------------------------------

We have seen that the existence and nature of Tolman temperature gradients depends \emph{both} on the spacetime in question, \emph{and} on the choice of 4-velocity for the heat bath of interest. For a heat bath that follows the trajectories of any timelike Killing vector one has
\begin{equation*}
A_a =  \nabla_a \ln ||K||= - \nabla_a \ln T;  \qquad T(x) = {T_0\over||K||}.
\end{equation*}
For a heat bath that follows the trajectories of a suitably chosen normal flow one has
\begin{equation*}
A_a =  \nabla_a \ln N= - \nabla_a \ln T;  \qquad T(x) = {T_0\over N}.
\end{equation*}
Sometimes these are the same (static spacetimes in block diagonal form), but typically  they are different. 

Also, given gravity's universality, the temperature gradient in thermal equilibrium states cannot depend on which material one is considering. In this way, the relation between fluid 4-acceleration and the equilibrium temperature gradient present in the system must be the same as that obtained for a photon gas in this paper, given by:
\begin{equation*}
A_a =  - \nabla_a \ln T. 
\end{equation*}
This general formula must work for all possible thermodynamically compatible 4-velocity fields.
By choosing a free-fall normal flow, the Tolman temperature gradient can be made to vanish. By choosing a normal flow compatible with Boyer--Lindquist coordinates, one can make a plausible definition of redshifted Hawking temperature that works all the way from the horizon to spatial infinity.

Finally it is worth emphasizing yet again that the existence of Tolman temperature gradients cannot be separated from some choice as to the 4-velocity of the heat bath. 
This has been implicit in many previous calculations~\cite{Israel:1976, Abreu:2010a, Abreu:2010b, Abreu:2010c, Abreu:2011, Padmanabhan:2003, Padmanabhan:2010, Padmanabhan:2017}, but it is worthwhile to make this point explicit.

%-----------------------------------------------------------------------------------------------------
\acknowledgements
%-----------------------------------------------------------------------------------------------------
JS was supported by a Victoria University of Wellington PhD Scholarship.
JS wishes to thank Cesar Uliana--Lima and Uli Zuelicke for helpful discussions.\\
MV was supported by the Marsden Fund, administered by the Royal Society of New Zealand. 
%-----------------------------------------------------------------------------------------------------

%-----------------------------------------------------------------------------------------------------

%-----------------------------------------------------------------------------------------------------

%-----------------------------------------------------------------------------------------------------

%-----------------------------------------------------------------------------------------------------

%-----------------------------------------------------------------------------------------------------
\end{document}